# Malicious code detection in android: the role of sequence characteristics and disassembling methods

Pinar G. Balikcioglu[1] · Melih Sirlanci[1,2] · Ozge A. Kucuk[1] · Bulut Ulukapi[1] · Ramazan K. Turkmen[1] · Cengiz Acarturk[1,3]

**Abstract**
The acceptance and widespread use of the Android operating system drew the attention of both legitimate developers and malware authors, which resulted in a significant number of benign and malicious applications available on various online markets. Since the signature-based methods fall short for detecting malicious software effectively considering the vast number of applications, machine learning techniques in this field have also become widespread. In this context, stating the acquired accuracy values in the contingency tables in malware detection studies has become a popular and efficient method and enabled researchers to evaluate their methodologies comparatively. In this study, we wanted to investigate and emphasize the factors that may affect the accuracy values of the models managed by researchers, particularly the disassembly method and the input data characteristics. Firstly, we developed a model that tackles the malware detection problem from a Natural Language Processing (NLP) perspective using Long Short-Term Memory (LSTM). Then, we experimented with different base units (instruction, basic block, method, and class) and representations of source code obtained from three commonly used disassembling tools (JEB, IDA, and Apktool) and examined the results. Our findings exhibit that the disassembly method and different input representations affect the model results. More specifically, the datasets collected by the Apktool achieved better results compared to the other two disassemblers.

**Keywords** Malware detection · LSTM · Natural language processing

## Abbreviations

| | |
|---|---|
| LSTM | Short-term long memory |
| NLP | Natural language processing |
| RNN | Recurrent neural network |
| DT | Decision trees |
| SVM | Support vector machines |
| NB | Naïve Bayes |
| LR | Logistic regression |
| EL | Ensemble learning |
| OL | Online learning |
| KNN | K-nearest neighbor |
| CNN | Convolutional neural network |
| MLP | Multilayer perceptron |
| ISM | Instruction as a sequence model |
| BSM | Blocks as sequences model |
| MSM | Methods as sequences model |
| CSM | Classes as sequences model |

✉ Cengiz Acarturk
cengiz.acarturk@uj.edu.pl

1 Cyber Security Department, Middle East Technical University, Ankara, Turkey
2 Computer Science and Engineering Department, Ohio State University, Columbus, OH, USA
3 Cognitive Science Department, Jagiellonian University, Krakow, Poland

## Introduction

Android mobile operating system has a market share of 73% for smartphones, and 43% for tablets [1]. Thanks to its open-source feature, the Android operating system offers an affordable and convenient interface for end-users. It also provides a well-established development environment and infrastructure for mobile software developers. Hence, an inevitable consequence is the close attention of malware authors, who target the Android platform by mostly embedding their malicious code into benign applications. Nowadays, malicious code detection, as implemented by antivirus software or anti-malware tools, is a native part of many desktop operating systems. Nevertheless, mobile operating systems are most vulnerable to malware. In 2020, Kaspersky Mobile products and technologies detected a 62.2% annual increase in the installation of malicious packages on Android environment [2]. This growing threat points out the urgent need for anti-malware tools running with high detection accuracies.

In general, malware analysis methods can be grouped under three different categories: static, dynamic, and hybrid analysis methods. The static analysis relies on examining the source code of a program and accompanying metadata without running the binary itself. The dynamic analysis aims to investigate the behavior of the executable by running it. Lastly, the hybrid analysis leverages the combinations of the other two analysis methods. Early anti-malware software employed signature-based protection mechanisms that created traces of known malware from existing malware recorded in constantly updated databases. In these systems, signatures, and traces are devised by utilizing the previously mentioned malware analysis techniques. The aforementioned paradigm suffered from the limitations of the size of the databases, the temporal gap between the detection of malware and the implementation of the signature in the database (cf. zero-day vulnerabilities), and the ease of obfuscation techniques creating versions of the same malware with different signatures. Therefore, malware detection methods that offer adaptive solutions, including machine learning techniques, have replaced signature-based methodologies. Unlike signature-based methods, the methods with machine learning (ML) techniques utilize malware analysis methodologies to perform feature extraction from Android application (apk) files [21], [39]–[42].

A common practice in evaluating ML-based malware detection methods and models, specifically the classification accuracy of binaries as benign or malware, is to conduct bench-marking by reporting accuracy values in contingency tables. This common practice has proved to be an efficient evaluation method, allowing researchers to perform comparative analyses of their methodologies. Nevertheless, besides the multi-parameter nature of learning machines such as deep learning methods, the other factors, like the input unit(as formed by the code structures), the input representation specified by the disassembling tool, and the length of sequences used in language modeling may also have a high impact on the results. However, due to the complex nature of malware detection, those factors and the parameter values relevant to them are usually conceived as operational assumptions rather than being treated as determining factors with a high impact on the results.

The goal of the present study is twofold: We present a methodology for Android malware detection using Long Short-Term Memory (LSTM) by following our previous research [37]. Then, we have a closer look at the factors that have an impact upon the model outcomes. In particular, we focus on the role of the disassembler software since it identifies the representation of the input. Our findings show that the disassembling method influences the processing, thus the model outcomes. Another factor is the form of the input data as specified by the researcher. In particular, a researcher may prefer to use a set of instructions, basic blocks, methods, or classes as the unit of input representation. Our findings reveal how different types of input representation influence the model outcomes.

For model development, we employed a specialized type of Recurrent Neural Network (RNN), namely the LSTM [3]. We handle malware detection from a Natural Language Processing (NLP) perspective. For this purpose, we develop and test the results in various basic units (instruction, basic block, method, and class) generated by the JEB Decompiler [4], the IDA Pro Disassembler [5], and the Apktool [6], which are widely used tools in reverse engineering apk files. More specifically, we reverse-engineered the apk files by the JEB Decompiler, the IDA Pro Disassembler, and the Apktool. First, the benign and malicious apk files were decoded, then the resulting codes were divided into instructions, basic blocks, methods, and classes. Finally, LSTM models were trained on those units with the Dalvik assembly code of malicious and benign apk files.

The following section describes the relevant studies about malware detection on Android by employing machine learning and deep learning techniques. Afterward, we present our proposed methodology, including the approach, the details on datasets, and the LSTM pipeline. We then report the results and present a discussion of the findings within the framework of the literature.

# Relevant work

The popularity of Android is of interest to malware developers as well as researchers who develop anti-malware solutions. Thus, in the past two decades, detecting and analyzing Android malware has become an active area of research in cyber security. An indispensable part of the research effort has been integrating machine learning techniques, more recently the adaption of deep learning methodologies for higher accuracy in detection rates, besides the opportunity to detect novel malware types. In this section, we present the relevant work on Android malware detection methods.

In a survey study, Sharma and Rattan conducted a review on Android malware detection [7]. They categorized malware analysis techniques into three groups, namely static, dynamic, and hybrid analysis. The static analysis refers to reviewing an apk file without executing the application, mainly focusing on the manifest and the source code file. In some instances, malware authors obfuscate the apk files to reduce the efficiency of malware detection by increasing the analysis time. Nevertheless, static analysis is a fast and popular technique for malware detection. An alternative to static malware analysis is dynamic analysis. With dynamic analysis, the application is executed, and its behavior is observed. In dynamic analysis, a separate, isolated environment is needed to run the application. Dynamic analysis is a more complex process that may become a resource- and time-consuming. However, it is a widely used and reliable technique. Lastly, hybrid strategies use the combination of static and dynamic analysis to inspect the application by utilizing the two analysis methods together. A significant advantage of static analysis is its lightweight infrastructure and low-cost efficiency. Therefore, in the present study, we used static analysis. We used smali files disassembled by alternative disassembly tools as inputs to the model.

A review of the relevant work shows that various machine learning models and model features have been employed in the static analysis of malicious code. The most common models and learning algorithms include Decision Trees (DT), Support Vector Machines (SVM), Naïve Bayes (NB), Logistic Regression (LR), Ensemble Learning (EL), Online Learning (OL), K-Nearest Neighbor (KNN), and neural networks (e.g., RNN, CNN, MLP, LSTM). In addition, various features have been selected for the models of Android malware detection, including application permissions, API calls, intents, opcode sequences, program graphs (e.g., function call graphs, control flow graphs), and hardware features obtained from Android apk files [8]–[19]. In our present study, we utilized a deep learning model, LSTM, and trained the model with the instruction sequences extracted from disassembled smali files of applications. In other words, we used opcode sequences as the features and LSTM as the learning model.

Numerous studies have employed Natural Language Processing (NLP) Deep Learning models to develop malware detection methods. For instance, Karbab et al. utilized NLP models and proposed a framework that relied on API method call sequences to detect and classify Android malware [24]. They obtained dex files from apk files for the API call extraction by using Python's zip library. The Dalvik assembly code was obtained from dex files with dexdump. The API call sequences were then extracted by utilizing regular expressions. Those sequences were used to generate semantic vectors with word2vec and then fed to a CNN (Convolutional Neural Network) model to perform detection and classification. The suggested framework achieved a detection accuracy of 96–99%. In a similar study [26], Ma et al. proposed a method called Droidetec, which represented applications in terms of natural language sequences by utilizing the API calls. They used the Androguard [43] tool to retrieve dex files from apk files. After extracting API sequences with some pre-processing, one-hot vector embedding and Skip-Gram model were used to set up the input vector of sequence. The detection was performed by forwarding this input to the LSTM-based detection model. Aside from malware detection, Droidetec adopted an attention mechanism that utilized weight distributions of APIs to localize the potentially malicious code and output a report providing specific information on malicious code segments. Droidetec reached over 97% accuracy for detection and a 91% hit rate for detecting malicious code. Wu et al. [25] employed a similar semantic view for malware detection. Instead of the API calls, the researchers used requested permissions extracted from the Android manifest file AndroidManifest.xml. After the extraction, a threshold mechanism was employed to filter out standard permissions. Then by word2vec, permissions were represented as a one-hot vector fed into the LSTM model for malware detection. A different approach was proposed by Xie et al., which offered an LSTM-based method that applied dynamic analysis [27]. They used system calls as the model features, and they represented the program's behavior with system-call sequences (viz. the behavioral language in their terminology). The model was tested with both host-based and Android-based datasets and achieved 99% and 95% accuracy, respectively. In another study, the authors used both static and dynamic analysis techniques giving an example of a hybrid method for malware analysis [33]. For the static analysis, they extracted permissions of applications from Android manifest files using the Apktool. The battery, binder, memory, and permissions were collected using emulators and feature collection platforms, then used as the features in the dynamic analysis. The researchers used various combinations of the features on RNN (Recurrent Neural Network) and LSTM models for comparative analysis. Their results showed that the LSTM outperformed RNN, achieving a detection rate of 97% for static analysis and 93% for dynamic analysis. More recently, a malware classification framework, namely ROCKY, performed malicious code detection by extracting API calls, permission, keywords, and tokens from decompiled source code instead of Dalvik bytecode or smali code [28]. It applied lexical analysis of the source code by utilizing sequences of N-tokens that include stop-tokens, feature-tokens, and long-tail tokens. The developers compared ROCKY with other neural network models reported in the previous work. Their findings revealed that ROCKY outperformed other models and features, achieving an accuracy of 97.5%. In contrast to the studies that applied semantic-based techniques using features like API calls and permissions, our models use the smali code of android applications to take advantage of the LSTM model in learning the behavior of the applications.

A further review of the application of the learning methods for malicious code detection reveals that opcode sequences have often been used in malware detection. For instance, McLaughlin et al. proposed a CNN-based malware detection approach [20]. They used the Baksmali [44] tool to extract opcode sequences and proposed a Hierarchical Denoising Network model to deal with the Gradient vanishing problem of LSTM by using encoders. So, the basic idea was to disassemble each apk file into smali files by using Baksmali, and then remove the operands to obtain opcode sequences from each method in all the smali files. Gathering the opcode sequences into a single opcode sequence makes it possible to represent the whole application. Text representations of applications were then fed to the CNN model to perform the detection. The main advantage of the study was that it did not require complex, manual preprocessing for data preparation. Parker et al. conducted a similar study using the same data preparation and disassembler tools [21]. However, they used two different classification methods, namely weighted sequence and Modified Multi-Layer Vector Space models. Their methods did not explicitly target Android malware detection. Their results achieved 79% accuracy. In another study, Li et al. adopted the same data preprocessing approach using a different disassembler, namely Apktool, to collect the smali files [22]. They used k-max pooling models and compared their model with a set of classical machine learning models, such as RF, DT, SVM, Naïve Bayes, and MLP. Their model reached 99% of accuracy and outperformed the others.

Numerous studies have employed features, such as opcode sequences or source code in LSTM models, alongside the previously mentioned research focusing on opcode sequences. For example, Deeprefiner was proposed as a semantic-based, two-layered method for malware detection [29]. Deeprefiner's first layer performs an initial detection by inputting vector representations of Android manifest files and the other XML files under the resources directory to an MLP model. Then, the unclassified applications are forwarded to the second detection layer for a more thorough inspection. In the second layer, dex bytecode, extracted by the Apktool, is represented as vectors using Skip-Gram modeling, then fed into the LSTM model. The preprocessing of the bytecode involves reducing all dex instructions to fifteen instruction categories without removing information like class names, fields, and methods.

Other studies that used opcode sequences with LSTM models include [31], in which the Apktool was used to obtain opcode sequences. The opcodes were then converted to 16-bit code representations. The model also utilized an opcode sequence encoder to reduce feature dimensions. In [34], Amin et al. adopted a similar approach using opcode sequences and LSTM for malware detection and family attribution. The authors conducted experiments with other neural network architectures. They stated that BiLSTM outperformed other models without the burden of manual preprocessing for the features. A similar study used opcode sequences retrieved from instruction call graphs [23]. The

**Table 1** The type and the number of malware samples used in Dataset 2

| Type | Number |
| --- | --- |
| Adware | 500 |
| Banking malware | 500 |
| SMS malware | 500 |
| Mobile riskware | 500 |

Androguard tool was used for decompiling the applications to obtain all the execution paths. The embedded opcode sequences were then fed into the LSTM model, and the detection was performed by evaluating the features of the graph representation alongside the LSTM. In [32], the researchers reported experiments with CNN and LSTM models applied on two datasets. For the LSTM model, the authors extracted Android bytecode using the Apktool. The bytes were then represented as vectors by one-hot encoding and then fed into the LSTM model. The proposed LSTM architecture achieved 70% accuracy for small datasets and 95.3% accuracy for large datasets. In our study, we used the instructions obtained from smali files without any manual preprocessing. We also included operands besides the opcodes in the instructions to avoid making simplification assumptions. A unique aspect of our study is the investigation of the role of the disassembler tool. We used Apktool, JEB, and IDA Pro to disassemble apk files to obtain smali files and reported the differences. The following section presents the methodology of the study.

## Methodology

### The dataset

We used two different datasets for the study, Dataset 1 and Dataset 2. Dataset 1 consisted of 600 benign and 787 malicious Android applications. We obtained the benign applications from the 2017 pack of the AndMal2017 dataset [35] and the malicious applications from the VirusShare website [36]. Since there was no information about the types of malware we collected from the Virusshare, we cannot state the types of malware we used in Dataset 1. On the other hand, Dataset 2 consisted of 700 benign and 2000 malicious Android applications. Both benign and malicious applications in Dataset 2 were collected from the AndMal2020 dataset [45]. The types and numbers of malware samples we used in Dataset 2 are shown in Table 1. Then, collected apk files were disassembled by JEB and IDA Pro, commercial tools, and the Apktool, an open-source software tool. Below, we describe the specific data extraction methods for the apk files in the datasets.

We used the JEB software tool demo version 3.17.1 [4] to obtain the static assembly output of the Android applications in the apk dataset. We made two modifications to the default JEB options: we removed the address information on each line and the annotations at the beginning of a disassembled output. The JEB command prompt is limited by the JEB disassembler running with a given Android application. So, we ran the JEB disassembler through the command prompt by loading the apk files. Then, we saved the assembly output of the corresponding apk files by selecting the Active View option in the Export menu through the user interface of the JEB disassembler.

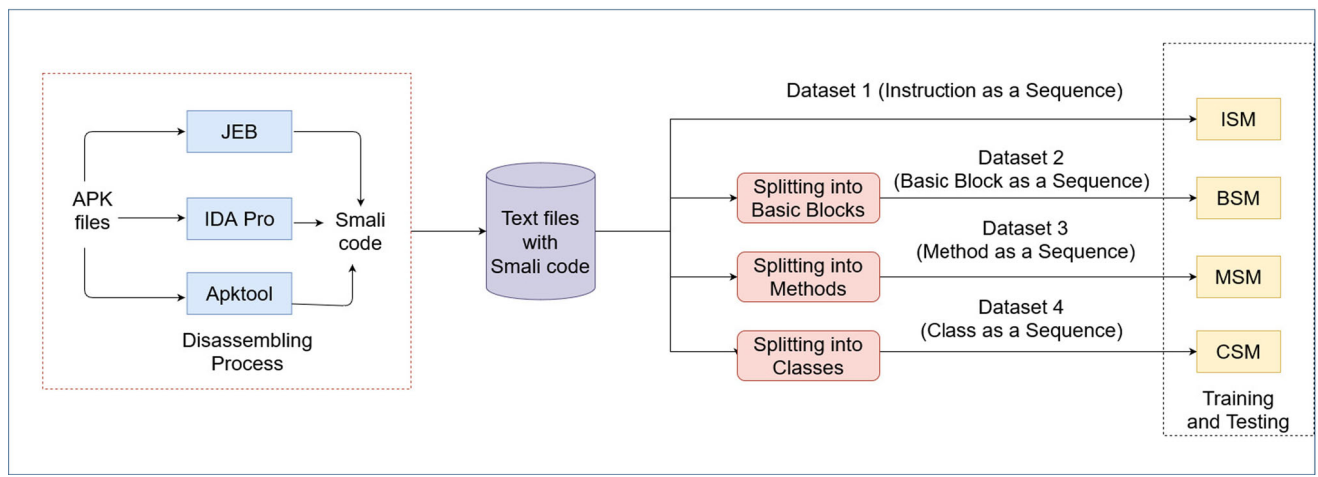

**Fig. 1** The modeling pipeline

IDA was the other software tool that we used to generate disassembly files [5]. Initially, we checked the decompilation output of IDA and noticed that the output includes additional unrelated information, which leads us to use IDAPython plugin provided by IDA. Basically, a python script traverses each function and saves the assembly instructions. We further used the command line features of IDA to automatically open each apk file, run the python script on the target apk, collect the instructions, and save them to a text file.

Our third disassembler Apktool, also used for generating the smali files [6]. The processing pipeline started with decoding the Android application by the tool. Then, the smali directory was recursively traversed. Finally, the decompiled application classes were collected into a single text file for each apk stored in separate smali files. Consequently, we generated a text file for each Android application, which included the disassembly output of the corresponding application.

Tables 2 and 3 present the number of sequences (assembly instructions) obtained through JEB, Apktool, and IDA and the number of files used to obtain the sequences for Dataset 1 and Dataset 2, respectively. The tables also present information about the number of sequences obtained in Basic Block, Method, and Class format and the number of files used, respectively.

Although we disassembled most Android applications using the three disassembler tools, the tools could not output the Dalvik assembly codes for a few applications. For example, out of 600 benign and 787 malicious Android applications in the Dataset 1, we obtained assembly outputs of 600 benign and 785 malicious applications on JEB, 600 benign and 787 malicious applications on Apktool, and 600 benign and 786 malicious applications on IDA Pro. Another challenge was that each disassembler generated the assembly code in a different size. We used similar number of sequences from the outputs of each disassembler to allow a comparative analysis by specifying a maximum value for the number of sequences used in training and testing. For instance, we limited our first model, namely the Instruction as a Sequence Model (ISM), to 200 million sequences (i.e., assembly instructions). Consequently, we obtained three sub-datasets in each Dataset (Dataset 1 and Dataset 2) by generating assembly output of the Android applications through three different tools. In the following section, we present the code structures of the Dalvik assembly code and the methods of implementation.

## The modeling pipeline

Android applications include several components (units) from a developer's perspective, such as instructions, basic blocks, methods, and classes. Those units provide the backbone of generating a language model by using Android application codes. In the present study, we focus on the units in the Dalvik assembly code obtained from Android applications, including instructions, basic blocks, methods, and classes. The overview of the modeling pipeline is presented in Fig. 1.

The instructions consist of opcode and operands, being the essential components of the assembly language. They have functionality and meaning interpretable by a machine, similar to a natural language sentence stating a proposition. The basic blocks are larger assembly language components that consist of a group of instructions that come up sequentially. So, basic blocks contain more complex functionality and meaning compared to instructions. Methods and classes are also two primary components of the Dalvik assembly code. A method includes at least one basic block, or usually more than a single basic block. So, methods are more complex than instructions and basic blocks. Classes include methods used to perform similar tasks, including more functionality and meaning than the constituents. Thus, classes are the largest semantic component of the assembly language.

Table 2 The properties of the dataset 1

| | Malicious | | | | | | Benign | | | | | |
|---|---|---|---|---|---|---|---|---|---|---|---|---|
| | Number of files | | | Number of sequences (million) | | | Number of files | | | Number of Sequences (million) | | |
| | JEB | Apk tool | IDA pro | JEB | Apk tool | IDA pro | JEB | Apk tool | IDA pro | JEB | Apk tool | IDA pro |
| Instruction | 738 | 585 | 706 | 200 | 200 | 200 | 598 | 409 | 600 | 200 | 200 | 200 |
| Basic block | 785 | 787 | 786 | 85 | 78 | 88 | 599 | 600 | 600 | 90 | 86 | 89 |
| Method | 785 | 787 | 786 | 12 | 10 | 12.5 | 600 | 588 | 600 | 12.5 | 13 | 12.5 |
| Class | 785 | 787 | 786 | 1.75 | 1.50 | 1.92 | 600 | 579 | 600 | 1.90 | 2 | 1.92 |

Table 3 The properties of the dataset 2

| | Malicious | | | | | | Benign | | | | | |
|---|---|---|---|---|---|---|---|---|---|---|---|---|
| | Number of files | | | Number of sequences (million) | | | Number of files | | | Number of sequences (million) | | |
| | JEB | Apk tool | IDA pro | JEB | Apk tool | IDA pro | JEB | Apk tool | IDA pro | JEB | Apk tool | IDA pro |
| Instruction | 1884 | 1760 | 1959 | 200 | 200 | 200 | 644 | 634 | 649 | 200 | 200 | 200 |
| Basic block | 1985 | 2000 | 1995 | 80 | 78 | 78 | 617 | 668 | 684 | 90 | 90 | 90 |
| Method | 1985 | 2000 | 1995 | 11.2 | 10.7 | 11.2 | 667 | 665 | 668 | 13 | 13 | 13 |
| Class | 1985 | 2000 | 1995 | 1.6 | 1.6 | 1.6 | 654 | 650 | 657 | 2 | 2 | 2 |

Upon building the dataset using JEB, IDA, and Apktool, we obtained text files, including the whole decompiled output of the corresponding apk files. By removing specific content, such as the comment lines beginning with "#", we obtained the dataset of instructions as sequences. Then, we put the sequential instructions together to create the second dataset of basic blocks as sequences by separating them from specific opcodes, such as "jump" and "call," since they create branches in the code. To create the third dataset, we separated the sequential instructions from the locations where the methods start and end. The third dataset included sequences consisting of methods in the assembly language, viz. methods as sequences. Lastly, we put the sequential instructions together for the fourth dataset by separating them from the locations where classes start. The fourth dataset included sequences consisting of classes in the assembly language, thus a dataset of classes as sequences

In the present study, we adapted the language modeling pipeline described in our previous study [37]. In the relevant work part, we mentioned that researchers use various ML models and algorithms. Most ML models require feature extraction and laborious manual preprocessing, which can be challenging. To avoid these difficulties and save up from time-consuming preprocessing, we focused on RNN models. RNN and its specialized architectures are the best options for problems, including sequential data in language modeling. Compared to other RNN architectures, the LSTM solves vanishing and exploding gradient issues better and achieves better results. Also, in our previous work [37], we worked with LSTM and achieved outstanding results. Therefore, the pipeline in this study was also built on the LSTM architecture. The architecture we proposed and used in the present study consisted of six sequential layers, namely Embedding, LSTM, Pooling, Dropout, and two Dense Layers at the end. Also, for each model, we used the same parameter values with [37]: the dropout rate = 0.2; the number of output nodes in LSTM layer = 64; the batch size = 128; learning rate = 0.0010000000474974513; the optimizer is Adam [also see 38 for selecting the parameter values]. A major difference from [37] was that we trained the models for five epochs in the present study since we observed that our models continued to learn for a few more epochs. Also, in addition to the ISM (Instruction as Sequences Model) and the BSM (Blocks as Sequences Model), we specified new sequence length values for our new models, namely MSM (Methods as Sequences Model) and CSM (Classes as Sequences Model). The values of the sequence length parameter were 15, 40, 500, and 2500 for ISM, BSM, MSM, and CMS, respectively.

In summary, in this section, we reported the code structures of the Dalvik assembly code, including instructions, basic blocks, methods, and classes. Next, we presented how we obtained those structures from the Dalvik assembly code outputs. Then, we introduced a summary of the language modeling pipeline used in the present study. We obtained 12 datasets based on three disassembly methods (JEB, IDA, and Apktool) and four models (ISM, BSM, MSM, and CSM).

# Results

In this section, we present the results of the LSTM models trained on the datasets. We used three datasets (JEB, IDA, Apktool) for Dataset 1 and Dataset 2 in our ISM experiments by simply removing the lines irrelevant to the program functionality in the assembly output, such as comment lines and the lines with blanks. We call this first set of datasets the base datasets since we did not change the data format. In the Basic Blocks Sequences Model (BSM) datasets, we performed preprocessing on the base datasets to change the data format. In preprocessing, we put the sequential instructions together by separating them from the basic block endpoints. So, we obtained the datasets that included the same data but in a different format. In the second set of datasets, each sequence consisted of a basic block. In the third set of datasets, we used the base datasets obtained from the three disassemblers and performed preprocessing to change the data format as intended. In the preprocessing phase, we put the instructions in the same methods together by separating them from the method's beginning and endpoints. Thus, we obtained a new set of datasets that included the same data with the ISM and BSM but in different formats. Those new datasets included sequences consisting of methods called MSM. In the last set of datasets, we worked on the base datasets by performing preprocessing to change the data format. In the preprocessing, we performed similar operations as in ISM and BSM by putting the sequential instructions in the same classes together by separating them from the class beginning points. In those datasets, each sequence consisted of a class and they are called CSM. We trained each model on JEB, IDA, and Apktool datasets using the aforementioned parameter values. Tables 4 and 5 present the results of the model evaluation for Dataset 1 and Dataset 2, respectively.

**Table 4** Evaluation results of the models for dataset 1 (TPR: True Positive rate, FPR: False Positive Rate, ACC: Accuracy)

|  | TPR (%) | | | FPR (%) | | | ACC (%) | | |
| --- | --- | --- | --- | --- | --- | --- | --- | --- | --- |
|  | JEB | Apk tool | IDA Pro | JEB | Apk tool | IDA Pro | JEB | Apk tool | IDA Pro |
| ISM | 67.0 | 74.7 | 69.4 | 32.2 | 42.1 | 33.3 | 67.8 | 66.8 | 68.5 |
| BSM | 62.2 | 71.5 | 66.3 | 18.7 | 18.2 | 20.8 | 72.0 | 76.7 | 72.8 |
| MSM | 70.6 | 80.6 | 69.7 | 15.0 | 6.10 | 12.4 | 77.9 | 87.4 | 78.7 |
| CSM | 72.1 | 81.6 | 75.2 | 11.0 | 4.90 | 10.8 | 81.0 | 88.7 | 82.2 |

**Table 5** Evaluation results of the models for dataset 2 (TPR: True Positive rate, FPR: False Positive Rate, ACC: Accuracy)

|  | TPR (%) | | | FPR (%) | | | ACC (%) | | |
| --- | --- | --- | --- | --- | --- | --- | --- | --- | --- |
|  | JEB | Apk tool | IDA Pro | JEB | Apk tool | IDA Pro | JEB | Apk tool | IDA Pro |
| ISM | 71.3 | 70.1 | 72.1 | 33.6 | 32.5 | 33.2 | 68.5 | 68.7 | 69.1 |
| BSM | 78.2 | 81.9 | 79.9 | 28.6 | 26.1 | 26.4 | 73.9 | 76.4 | 75.7 |
| MSM | 85.0 | 88.8 | 85.0 | 22.7 | 19.2 | 22.2 | 80.4 | 83.8 | 80.9 |
| CSM | 86.8 | 91.4 | 89.7 | 18.6 | 16.4 | 18.5 | 84.2 | 86.7 | 86.1 |

## Evaluation results for dataset 1

While using Dataset 1, we reached the highest True Positive rate (TPR) of 74,7% for the ISM model when we trained the model with the code disassembled by Apktool. The outputs of JEB and IDA Pro achieved TPR of 67.0% and 69.4%, respectively, in the ISM model. After switching the model to the BSM, TPR for all the tools decreased by around 3-4%. Nonetheless, Apktool achieved the highest TPR of 71.5% also in the BSM model. When the MSM model is used, the TPR of Apktool and JEB increased to 80.6% and 70.6%, respectively. The TPR of IDA Pro also increased by around 3%, which was a smaller increase compared to the other two disassemblers. Lastly, we observed the highest TPR for all disassemblers while using the CSM model. In the CSM model, Apktool, JEB, and IDA Pro reached the TPR of 81.6%, 72.1%, and 75.2%, respectively.

On the other hand, when JEB outputs are used, the ISM model has the minimum False Positive rate (FPR) of 32,2%. In the ISM model, the FPR of Apktool and IDA Pro are 42.1% and 33.3%, respectively. When we move to the BSM model, the FPR rates of all disassemblers drop significantly. Apktool achieved the lowest FPR of 18.2% in the BSM model. The other two disassemblers, JEB and IDA Pro, reached the FPR of 18.7% and 20.8% in the BSM model. We observed a continuous reduction in the FPR as we went up to the CSM from the ISM. The lowest FPR we observed in the MSM was Apktool with 6.1%. Apktool also achieved the lowest FPR of 4.9% in the CSM model. As we observed in the TPR rates, FPR rates were minimum when we used the CSM model for all three disassemblers.

We observed similar patterns and results for the ACC values using Dataset 1. In the ISM model, ACC values of disassemblers were close to each other, while the highest of the three was IDA Pro with 68.5%. The ACC values of JEB and IDA Pro increased by around 4-5% whenever we switched the model from the ISM to the CSM. However, Apktool achieved the highest ACC values of 76.7%, 87.4%, and 88.7%, respectively, in the BSM, the MSM, and the CSM models.

## Evaluation results for dataset 2

When we switched to Dataset 2, we observed some improvements and reductions in the TPR, the FPR, and the ACC values. In the ISM model, the TPR values do not change drastically among the disassemblers. IDA Pro reached the highest TPR of 72.1% in the ISM model. We observed continuous improvements in the TPR values as we went up through the CSM model. All TPR values increased by around 10-15% compared to Dataset 1. Apktool reached the highest TPR values in the BSM, the MSM, and the CSM models with 81.9%, 88.8%, and 91.4%. Also, JEB and IDA Pro achieved a TPR of 86.8% and 89.7% in the CSM model.

When we look into the FPR values, we also saw an increase in around 8-10% for all the values. In the ISM, FPR values are close, the lowest being Apktool with an FPR of 32.5%. Again, we observed a similar pattern and decrease in the FPR values as we went up through the CSM model. The FPR values do not change drastically among disassemblers in all models. However, Apktool also reached the lowest FPR values in the BSM, the MSM, and the CSM models. The best FPR value is 16.4% and achieved with Apktool using the CSM model.

We also observed similar patterns and trends for the ACC values using Dataset 2. In the ISM model, ACC values of disassemblers were close, while the highest of the three was IDA Pro with 69.1%. In all the other models (the BSM, the MSM, and the CSM), Apktool reached the highest ACC values of 76.4%, 83.8%, and 86.7%, respectively. However, we observed a decrease in Apktool's ACC values compared to Dataset 1. On the other hand, ACC values of JEB and IDA Pro slightly increased when we switched to Dataset 2. They achieved their highest values in the CSM model. JEB reached an 84.2% accuracy, whereas IDA Pro achieved 86.1%, which is significantly close to the ACC value of Apktool.

## Discussion

In the present study, we investigated two factors that impact the ML-based model outputs in classifying benign and malicious code pieces. The first one is the method of parsing the data into sequences, which shapes the sequence characteristics in the train set and the test set. The second is the methods of disassembling the code pieces, which eventually generate data with variation in input representation. This section presents a discussion of the findings on those two dimensions.

### Sequence characteristics

A significant requirement for using language modeling techniques is the assembly data as input sequences. Since the disassembly instructions already form sequences, the model design is straightforward. The assembly instructions are central to an Android application that can be reached easily through decompilation or decoding processes if the source code is not accessible. Assembly instructions reflect the functionality of an Android application, including the grammatical components that show that the assembly instructions include meaningful information and patterns. Nevertheless, the characteristics of the underlying infrastructure of the input representation have an impact on the performance. A significant dimension of input characteristics is the unit of the data. Besides the instructions, assembly code includes structures essential from a developer's perspective, such as basic block, method, and class. A basic block is the smallest unit that consists of more than one assembly instruction that is functionally related. A method includes more than one basic block, so the method contains more assembly instructions that are functionally related to performing a more extensive function than a basic block. On the other hand, a class includes several methods together for a bigger purpose. Tables 6 and 7 show the average number of tokens in our models and the accuracy values reported in Tables 4 and 5, repeated for comparison.

The findings reveal that the accuracy increases as we move from the instruction unit to the class unit. There can be two possible explanations for that, related to each other. As we go from instruction to class, we take assembly instructions in semantic relationships. The semantic relations lead the data to have patterns that allow generating language models. The other explanation is the increase in sequence length while moving from instruction unit to class unit. As shown in Tables 6 and 7, the average number of tokens in the units increases from ISM to CSM, in line with increased accuracy. So, the model accuracy increases in line with the increase in sequence length. It is feasible since the increase in sequence length helps the model cover more tokens in one sequence, leading to better learning from the data.

**Table 6** Average number of tokens and accuracy values for the models using Dataset 1

|     | Average token |         |         | ACC (%) |         |         |
|-----|---------------|---------|---------|---------|---------|---------|
|     | JEB           | Apktool | IDA pro | JEB     | Apktool | IDA pro |
| ISM | 5.44          | 5.52    | 5.97    | 67.8    | 66.8    | 68.5    |
| BSM | 12.2          | 19      | 14.9    | 72.0    | 76.7    | 72.8    |
| MSM | 71.1          | 117     | 81.2    | 77.9    | 87.4    | 78.7    |
| CSM | 450           | 676     | 505     | 81.0    | 88.7    | 82.2    |

**Table 7** Average number of tokens and accuracy values for the models using dataset 2

|     | Average token |         |         | ACC (%) |         |         |
|-----|---------------|---------|---------|---------|---------|---------|
|     | JEB           | Apktool | IDA pro | JEB     | Apktool | IDA pro |
| ISM | 5.49          | 7.23    | 6.50    | 68.5    | 68.7    | 69.1    |
| BSM | 12.32         | 17.92   | 14.95   | 73.9    | 76.4    | 75.7    |
| MSM | 72.81         | 103.33  | 82.52   | 80.4    | 83.8    | 80.9    |
| CSM | 457.95        | 624.74  | 515.88  | 84.2    | 86.7    | 86.1    |

### The methods of disassembling

We built the datasets using three disassemblers, namely JEB, IDA, and Apktool, to disassemble the same apk files. The tools generated data with syntactic differences due to their different method of disassembling. Those differences in input representations resulted in different accuracy values, as reported in Tables 4 and 5. The results obtained through testing the models and some statistical information about the datasets used in the training and testing process are shown in Tables 6 and 7.

The results revealed a minor difference between JEB and IDA in the accuracy values. The output of IDA included a slightly larger number of tokens in sequences, leading to slightly better accuracy than JEB. As discussed in the previous section, longer sequences cover more data, leading to better learning. So, the number of tokens might have caused the slight accuracy difference between IDA and JEB. On the other hand, the data collected from the Apktool revealed significant differences in accuracy compared to the others. The output of the Apktool includes the path of the corresponding method besides the method's name. Table 8 presents sample instructions from each data source. The table consists of two lines of codes (instructions) from each of the disassembling tools. We present an example of the "invoke" command since the difference among the disassemblers is mainly observed in that command.

**Table 8** Sample instructions from each data source

| | |
|---|---|
| JEB | invoke-static AppEntry->access$100()AppEntry |
| JEB | invoke-static InstallOfferPingUtils->PingAndExit(Activity, String, Z, Z, Z)V, v0, v1, v3, v3, v2 |
| IDA Pro | invoke-static {}, <ref AppEntry.access100()$AppEntry_{a}ccess$100@L> |
| IDA Pro | invoke-static {v0, v1, v3, v3, v2}, <void InstallOfferPingUtils.PingAndExit(ref, ref, boolean, boolean, boolean) InstallOfferPingUtils_PingAndExit@VLLZZZ> |
| Apktool | invoke-static {}, Lair/com/adobe/connectpro/AppEntry;->access$100()Lair/com/adobe/connectpro/AppEntry; |
| Apktool | invoke-static {v0, v1, v3, v3, v2}, Lcom/adobe/air/InstallOfferPingUtils;->PingAndExit(Landroid/app/Activity;Ljava/lang/String;ZZZ)V |

The presence of the path information likely facilitated the classification of the data as malicious and benign. Moreover, the Apktool data consists of longer sequences due to the path information included in the Dalvik assembly code output. Those characteristics of the Apktool might have impacted the accuracy rates positively. Accordingly, the path information might be an indispensable factor that helps to represent the characteristics of an assembly code coming from one of the classes, malicious or benign.

Although a complete comparison to the available work in the literature is not possible due to differences among technical infrastructures, below, we present a partial comparison of the findings in the literature. Most of the studies which employed language modeling for malware detection used the Dalvik assembly code from Android applications. For example, in [29], [31], and [32], Apktool was used to collect the assembly data. In [30], Baksmali was used, whereas in [33], Androguard was used. In the present study, we proposed that the input data characteristics are an important factor that might affect the model accuracy. Our findings revealed that the choice of the disassembler requires more careful attention in malicious code detection. Most previous studies present one side of the coin by making operational assumptions in parameter values and preprocessing the input data, thus presenting a partial view of a larger picture. For instance, in [30], [31], and [33], researchers investigated the effect of the sequence length parameter on the success of the LSTM language models. While [30] and [31] focused on the data consisting of opcode sequences, and [33] worked on the data, including system call sequences. The accuracy rates improved by increasing the value of the sequence length parameter up to a certain point, suggesting that the sequence length may be influenced by other factors, such as the structure of the input units in assembly code.

## Conclusion

In the past decade, malware detection has gained significant importance due to the rapid growth of information and communication technologies (ICT). The development of polymorphic and metamorphic malware has brought the need for developing novel antimalware methods since traditional detection methods have lost their effectiveness against continuously changing features of the malicious content and zero-day attacks. Deep neural networks have been proposed as a good candidate for malware detection since they provide adaptability through learning.

We investigated Android assembly code obtained through three disassembler tools in the present study: JEB, IDA, and Apktool. The assembly data obtained from those tools showed differences in their representation as input data, and those differences led to minor differences in model accuracy between JEB and IDA data. We also investigated the sequence unit in the input data (viz. instruction, basic block, method, and class). Overall, the datasets obtained by the Apktool revealed better results compared to the other two disassemblers. This performance improvement in the Apktool dataset may be explained by the presence of the path information of each function and class in the Apktool. Another likely reason is that the sequences in the Apktool dataset are considerably longer than the sequences in the other two datasets; this gave Apktool an advantage in accuracy over JEB and IDA. Future research should clarify the source of the higher accuracy in the Apktool over the others since, in their recent form, our datasets and the models do not present a clear distinction among the dataset characteristics. Also, an investigation of the overall size of the datasets on the accuracy values, besides the characteristics of the input data should be carried out. Moreover, for the RNN model implementation we employed LSTM. There is also another architecture called Gated Recurrent Unit cell (GRU). Even GRU is said to have less expressibility than LSTM in some cases, it may have some advantages such as faster training. As a future work, a comparison study between LSTM and GRU should also be carried out to see the effects on accuracy values.

**Acknowledgements** Not applicable.

**Funding** Not applicable.

**Availability of data and materials** We used open datasets such as the 2017 pack of the AndMal2017 dataset [35], AndMal2020 dataset [45], and the VirusShare website [36] in our study. Additionally, we utilized android apk decompiler tools such as JEB and IDA Pro, commercial tools, and the Apktool, an open-source software tool.

## Declarations